\documentclass[aps,pra,reprint,superscriptaddress,floatfix]{revtex4-1}
\usepackage[colorlinks=true]{hyperref}

\usepackage{graphicx}
\usepackage[normalem]{ulem}
\usepackage{amsmath,amstext,amssymb}
\usepackage{dcolumn}
\usepackage{bm}

\newcommand{\Ef}{{\mathcal E}}
\newcommand{\imm}{~{\rm mm^{-1}}}
\newcommand{\icm}{~{\rm cm^{-1}}}

\newcommand{\mic}{~\mu{\rm m}}

\begin{document}
\sloppy

\title{Review of measurements of Kerr nonlinearities in lithium niobate: the role of the delayed Raman response }

\author{Morten Bache}\email{moba@fotonik.dtu.dk}
\affiliation{DTU Fotonik, Department of Photonics Engineering, Technical University of Denmark, DK-2800 Kgs. Lyngby, Denmark}%
\author{Roland Schiek}
\affiliation{Hochschule Regensburg, Pr¨ufeninger Strasse 58, 93049 Regensburg, Germany}%
\date{\today}

\begin{abstract}
    We review measurements in the literature regarding Kerr nonlinearity of lithium niobate. Of particular interest is to separate the value of the electronic Kerr nonlinearity and the strength of the delayed Raman nonlinearity. The spectral shape of the Raman response in LN is discussed and we restrict ourselves to the case where the pump is polarized along the optical axis. This configuration is important for so-called type 0 noncritical interaction, which is used in QPM devices and in recent supercontinuum generation and few-cycle soliton compression experiments.
\end{abstract}

\maketitle

\section{Introduction}

Lithium niobate (LiNbO$_3$, LN) is one of the earliest frequency conversion crystals that was used, and its optical linear and nonlinear properties have been studied in a large number of publications over the past 5 decades. It remains today one of the most important near-IR nonlinear crystals, so it is striking that its cubic nonlinear properties still need to be revised and investigated, as they play a crucial role in recent ultrafast experiments \cite{Langrock:2007,Phillips:2011,zhou:2012}. In fact, despite very early and thorough studies of the incredibly complex and rich nature of the cubic nonlinearity, where the Raman spectra were measured in detail \cite{barker:1967,Kaminow:1967,Johnston:1968}, still many key parameters needed for modeling and understanding ultrafast interaction are unknown. The purpose of this review is to present various measurements of the Kerr nonlinearity in LN and through the analysis try to propose good parameters usable of modeling the instantaneous (electronic) and delayed (Raman) nonlinearity in LN. We focus on the specific case of so-called type 0 interaction, where both the pump and the generated second-harmonic are polarized along the optical axis; this interaction is widely used in the nonlinear optical community as it exploits the largest $d_{33}$ component of the quadratic nonlinearity, and it is in particular used for devices based on quasi-phase matching (QPM) interaction.

An initial motivation for carrying out a literature study was a recent experiment by one of us:
In Schiek et al. \cite{Schiek:2005} four-wave mixing (FWM) was used to measure the Raman response of LN. A strong pump at $\lambda_p=1.064\mic$ and a weaker signal varied between $\lambda_s=1.07-1.25\mic$ were overlapped in the LN sample. This generated a new photon through FWM with the frequency $2\omega_p-\omega_s$. The LN was $X$-cut, and the pump and probe were both polarized along the $Z$-axis, which contains the largest quadratic nonlinear tensor component $d_{33}$. This corresponds to $\theta=\pi/2$, so the beams have extraordinary polarization. By analyzing the FWM signal, it could be broken down to $\chi^{(3)}=\chi_{E}^{(3)}+\chi_{\rm casc}^{(3)}+\chi_{R}^{(3)}+\chi_{\rm IR}^{(3)}$. Thus, contributions to the total nonlinearity came from the electronic Kerr effect $\chi_{E}^{(3)}$, cascaded quadratic nonlinearities $\chi_{\rm casc}^{(3)}$, the Raman effect $\chi_{R}^{(3)}$, and finally IR cascading $\chi_{\rm IR}^{(3)}$. By using well-known formulas for cascaded quadratic nonlinearities, tabulated Raman data from the literature, and IR dispersion relations, the electronic component could be extracted. The surprising result of this study was that it found a very large electronic Kerr nonlinearity for the $ZZZZ$ component
\begin{align}
    \label{eq:chi3-E-Schiek}
    \chi_E^{(3)}&=0.56\times 10^{4}~{\rm pm^2/V^2}\\
    \label{eq:n2-E-Schiek}
    n^I_{2,E}&=34.6\times 10^{-20}~{\rm m^2/W}
\end{align}
When modeling femtosecond pulse propagation the "total" Kerr nonlinearity is usually given as the sum of electronic and the Raman term, and in this case it gave
\begin{align}
    \label{eq:chi3-total-Schiek}
    \chi_{\rm tot}^{(3)}&=1.4\times 10^{4}~{\rm pm^2/V^2}\\
    \label{eq:n2-total-Schiek}
    n^I_{2,\rm tot}&=83.3\times 10^{-20}~{\rm m^2/W}
\end{align}
and we can therefore calculate the overall Raman fraction as
\begin{align}\label{eq:fR-Schiek}
    f_R&=\chi_{R}^{(3)}/\chi_{\rm tot}^{(3)}=0.58
\end{align}
which is a very important number to know when implementing the Raman effect in a numerical simulation. 
Finally the following ratio was found 
\begin{align}\label{eq:DC-ratio-Schiek}
 \chi_{R}^{(3)}/{\rm max}[\chi_{R,\sigma}^{(3)}({\rm peak})]=&0.165 
\end{align}
This represents the ratio between the DC value and the peak of all the Raman lines (i.e. the value at zero frequency when the Raman spectrum is normalized to have its peak at unity).

These numbers are much larger than what the literature have found so far: typically Z-scan measurement revealed near-IR values of $n^I_2\simeq 10\times 10^{-20}~{\rm m^2/W}$ for this tensor component \cite{desalvo:1996}, and although it was not specifically stated there this value is the sum of electronic and Raman nonlinear contributions (see also review in \cite{bache:2010}). Moreover,  consensus from the fiber optics community says that Raman fractions should be on the order $f_R=0.1-0.2$. The question is whether LN literature data can confirm the surprisingly large values found by Schiek et al.

\section{Raman term}

In the standard definition of a vibrational Raman-mode, it is described in time domain as a delayed fractional response adding to the instantaneous electronic cubic nonlinearity. Consider the nonlinear susceptibility \cite{agrawal:2007}
\begin{multline}\label{eq:chi3-el-Raman-time}
    \chi^{(3)}(t-t_1,t-t_2,t-t_3)=\\
    \chi_{\rm tot}^{(3)}R(t-t_1)\delta(t-t_2)\delta(t-t_3)
\end{multline}
where $\chi_{\rm tot}^{(3)}$ is the "total" nonlinearity; as we see below it is the sum of the electronic and DC Raman nonlinearity. The response function $R$ is defined as to include contributions from both electronic origin and from the vibrational Raman effect from a number of Raman modes
\begin{align}\label{eq:Rt}
    R(t)=(1-f_R)\delta(t)+f_R h_R(t)
\end{align}
where $f_R$ is the fractional contribution of the Raman response to the total nonlinear polarization, and
\begin{align}\label{eq:hRt}
    h_R(t)\equiv \sum_\sigma f_{R,\sigma}h_{R,\sigma}(t)
\end{align}
where $f_{R,\sigma}$ is the relative strength of mode $\sigma$ so $\sum_\sigma f_{R,\sigma}=1$. $R$ is normalized so $\int_{-\infty}^\infty dt R(t)=1$, and thus $h_R$ is normalized as well.

In frequency domain we have equivalently
\begin{align}\label{eq:chi3-el-Raman-frequency}
    \chi^{(3)}(\Omega)&=\chi_{E}^{(3)}+\chi_{R}^{(3)}(\Omega)
\end{align}
where
\begin{align}\label{eq:chi3-Raman}
    \chi_{R}^{(3)}(\Omega)=\chi_{\rm tot}^{(3)}f_R \sum_{\sigma} f_{R,\sigma}h_{R,\sigma}(\Omega)
\end{align}
and $h_{R,\sigma}(\Omega)$ is the Fourier transform of $h_{R,\sigma}(t)$. This definition means that the DC value of the Kerr nonlinear coefficient is \begin{align}\label{eq:chi3-zscan}
\chi_{\rm tot}^{(3)}=\chi_{\rm E}^{(3)} + \chi_{R}^{(3)}(\Omega=0)
\end{align}
This is the value most often measured, since standard $Z$-scan measurements have been performed with multi-ps pulses (see overview in Ref. \cite{bache:2010}).

The standard model for $h_R$ is a complex Lorentzian in frequency domain
\begin{align}\label{eq:hR-omega}
    h_{R,\sigma}(\Omega)&=\frac{\omega_\sigma^2}{\omega_\sigma^2 +i \Gamma_\sigma \Omega-\Omega^2}
\end{align}
whose normalized time form is \cite{blow:1989}
\begin{align}\label{eq:hR}
    h_{R,\sigma}(t)=\frac{\tau_{1,\sigma}^2+\tau_{2,\sigma}^2} {\tau_{1,\sigma}\tau_{2,\sigma}^2}\exp(-t/\tau_{2,\sigma})\sin(t/\tau_{1,\sigma}), \quad t>0
\end{align}
where $\tau_{1,\sigma}$ is the period of the temporal oscillations of mode $\sigma$ and $\tau_{2,\sigma}$ is characterizes its decay. Note that the response function is only defined for $t>0$ for reasons of causality. Taking the Fourier transform we get in frequency domain
\begin{align}\label{eq:hRw}
    h_{R,\sigma}(\Omega)&=\int_{-\infty}^\infty dt e^{-i\Omega t} h_{R,\sigma}(t)
    \nonumber\\
    &=\frac{\tau_{1,\sigma}^2+\tau_{2,\sigma}^2} {\tau_{2,\sigma}^2-\tau_{1,\sigma}^2(\Omega \tau_{2,\sigma} -i)^2}
\end{align}
The normalization is checked by noting that $h_{R,\sigma}(\Omega=0)=1$  and $h_{R,\sigma}(\Omega=0)=\int_{-\infty}^\infty dt h_{R,\sigma}(t)$. Now let us rewrite the part of the Raman response of Eq. (\ref{eq:hR-omega}) that contributes to the frequency variation, which we conveniently normalize by considering that at $\Omega=0$ it should be unity
\begin{align}\label{eq:hRw-Schiek}
    \frac{\omega_\sigma^2}{\omega_\sigma^2 +i \Gamma_\sigma \Omega-\Omega^2}&\equiv
    \frac{\Omega_{1,\sigma}^2+\Omega_{2,\sigma}^2}{\Omega_{1,\sigma}^2- (\Omega-i \Omega_{2,\sigma})^2}\\
    &\equiv \frac{\tau_{1,\sigma}^2+\tau_{2,\sigma}^2}{\tau_{2,\sigma}^2-\tau_{1,\sigma}^2(\Omega \tau_{2,\sigma} -i)^2}
\end{align}
where in the first line we have introduced
\begin{align}\label{eq:Omega2}
    \Omega_{2,\sigma}&\equiv \Gamma_\sigma/2\\
    \label{eq:Omega1}
    \Omega_{1,\sigma}&\equiv\sqrt{\omega_\sigma^2-\Gamma_\sigma^2/4}
    =\sqrt{\omega_\sigma^2-\Omega_{2,\sigma}^2}
\end{align}
and now $\Omega_{2,\sigma}$ is the HWHM linewidth. In the second line we used
\begin{align}
    \label{eq:tau1}
    \tau_{1,\sigma}&\equiv \Omega_{1,\sigma}^{-1}=[\omega_\sigma^2-\Gamma_\sigma^2/4]^{-1/2}\\
    \label{eq:tau2}
    \tau_{2,\sigma}&\equiv \Omega_{2,\sigma}^{-1}=2/\Gamma_\sigma
\end{align}
and we now have the desired form of Eq. (\ref{eq:hR-omega}).

\section{Raman parameters and Raman gain measurements of Johnston and Kaminow}
\label{sec:JK}

Early studies by Johnston and Kaminow \cite{Kaminow:1967} found four Raman lines, and even measured an absolute scattering strength for each mode. Using these data we can reconstruct the absolute values of the Raman peaks in mks units using the equations of \cite{Johnston:1968}. The only uncertainty is that they are unclear about the reported linewidths (whether what they report is the FWHM or not), and compared to other studies in particular \cite{Kaminow:1967} seems to measure quite narrow linewidths, whereas \cite{Johnston:1968} only measured the two dominant lines (at different temperatures) with results more in the trend of other studies. In Table I of \cite{Kaminow:1967} the transverse "T" modes of the type 0 interaction (all waves polarized along $Z$) are reported with scattering efficiencies $S_{33}/ld\Omega$ in units $\icm {\rm sr}^{-1}$. The linewidths are reported as $2\gamma_\sigma$, which we assume to be FWHM (i.e. $\gamma$ is the HWHM, and the notation $\gamma$ is used to emphasize that it is different from the $\Gamma$ used above: $\gamma=\Gamma/2$), however the linewidths of $2\gamma_\sigma=15$, 8, 6 and $15\icm$ for the four modes located at $253$, 275, 334 and $637\icm$, respectively, seem quite low. Not unsurprisingly, they later correct these linewidths, stating that they were 30\% too small, see \cite[p. 3497]{Johnston:1970} and \cite[p. 1049]{Johnston:1968}.
Proceeding with specific calculations, Eq. (7) in \cite{Johnston:1968} can be used to find the power-gain coefficient $g_s$ relative to the pump intensity $I_p$ (note the corrected equation from the erratum \cite{Kaminow:1969} is used)
\begin{align}\label{eq:GoI}
    \frac{g_s}{I_p}=\frac{16\pi^2 c^2}{\hbar \omega_{s}^3}\frac{S_\sigma/ld\Omega}{n_{s}^2(n_{0,\sigma}+1)2\gamma_\sigma}
\end{align}
where (all in mks units) $\omega_{s}$ is the absolute frequency of the Stokes wave, $n_p$ and $n_s$ are the linear refractive indices of the pump and Stokes waves, $n_{0,\sigma}=1/[\exp(\hbar \omega_\sigma/k_B T)-1]$ is the Bose population number of the Raman frequency at the temperature $T$. Finally, $S_\sigma/ld\Omega$ is the scattering efficiency (again the corrected equation from the erratum \cite{Kaminow:1969} is used)
\begin{align}\label{eq:scattering-cross-JK}
    S_\sigma/ld\Omega=\frac{(n_{0,\sigma}+1)\hbar\omega_s^4n_s}{32\pi^2\varepsilon_0^2 c^4\rho_\sigma\omega_\sigma n_p}|(\partial\bar\alpha/\partial q)_0|^2
\end{align}
where $\rho_\sigma$ is an effective reduced mass per unit volume of the mode. An important parameter $(\partial\bar\alpha/\partial q)_0$ is introduced, which is the derivative of the optical polarizability with respect to the normal coordinate $q$ of the vibration. Below we discuss this in further detail and show how it is connected with the first order transition hyperpolarizability. Note also the well-known scaling with $\omega_s^4$.

\begin{figure}[tb]
\includegraphics[width=8.5cm]{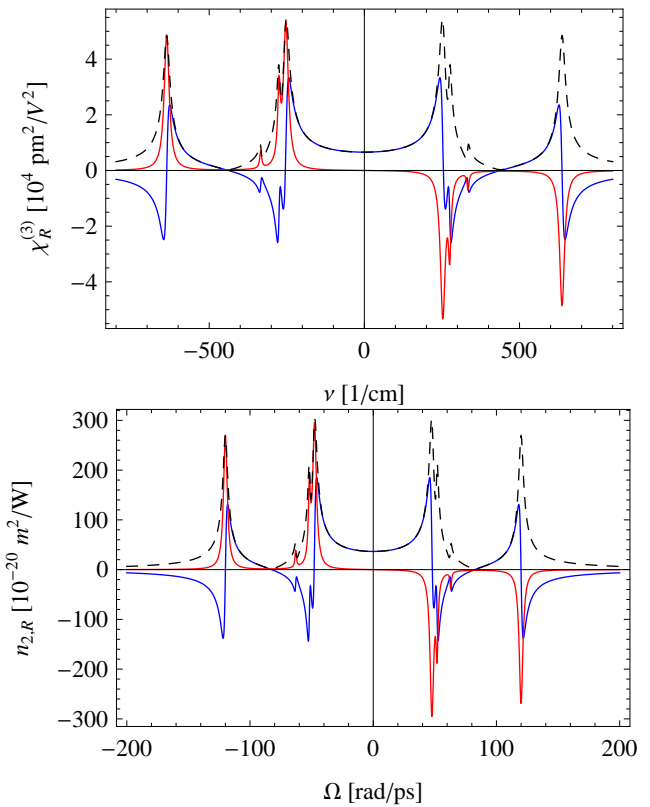}
\caption{\label{fig:Raman-KJ} Data from Kaminow and Johnston \cite{Kaminow:1967} measured for $\lambda=0.488\mic$. The Raman response $\chi^{(3)}_R$ vs. inverse wavenumber (top) and the corresponding Raman contribution to the nonlinear refractive index $n^I_{2,R}$ vs. angular frequency (bottom). The blue curve is the real (dispersive) part while the red curve is the imaginary (absorptive) part. The dashed black line indicates the absolute value. Note that the linewidths have been increased 30\% to the values of \cite{Kaminow:1967}, see more in \cite[p. 3497]{Johnston:1970} and \cite[p. 1049]{Johnston:1968}, and that the gain calculations use the corrected equations from the erratum \cite{Kaminow:1969}.
    }
\end{figure}

The experiment was conducted at room temperature and with $\lambda_p=0.488\mic$. We now calculate the peak gain values of the four modes, which gives \cite{Kaminow:1967,Kaminow:1969}
\begin{align}\label{eq:GoI-JK}
    \frac{g_s}{I_p}&=\{14.7, 7.2, 1.7, 13.5\}~{\rm cm/GW}
\\\nonumber &\text{(
using 30\% larger $\gamma_\sigma$)}
\end{align}
Now we can relate this Raman gain to a nonlinear susceptibility by using that the polarization response of a Kerr interaction in the type 0 case is [Eq. (B8) in \cite{bache:2010}]
\begin{align}\label{eq:PNL3}
P_{\rm NL}^{(3)}=\frac{3}{4}\varepsilon_0 [\chi^{(3)}_{{\rm eff},ii}|\Ef_{i}|^2+2\chi_{{\rm eff},ij}^{(3)}|\Ef_{j}|^2]\Ef_i
\end{align}
where $\chi^{(3)}_{{\rm eff},ii}$ is the effective nonlinear tensor component that depends on the birefringent properties of the crystal; for type 0 we have $\chi^{(3)}_{{\rm eff},ii}=\chi^{(3)}_{{\rm eff},ij}\equiv \chi^{(3)}$. The connection to the Kerr nonlinear index is 
\begin{align}\label{eq:n2-tilde}
    n^I_2=\frac{3\chi^{(3)}}{4\varepsilon_0c n^2}
\end{align}
where by definition $n=n_0+n_2^I I$, i.e. a nonlinear index change that is proportional to the intensity. From the SVEA we get that the cw wave equation for the Stokes field $\Ef_s$ at the frequency $\omega_s=\omega_p-\omega_\sigma$ is
\begin{align}\label{eq:svea-we}
    2ik_s \frac{d\Ef_s}{dz}=-\frac{\omega_s^2}{\varepsilon_0c^2}P_{\rm NL}^{(3)}(\omega_s)
\end{align}
where $k_s=n_s\omega_s/c$. The Stokes wave sees Raman gain from the XPM of the pump, and under the approximation of a weak gain the Stokes SPM is neglected. Thus, neglecting SPM 
\begin{align}\label{eq:svea-we-Stokes-gain}
    \frac{d\Ef_s}{dz}\simeq i\frac{3\omega_s}{4n_sc}\chi^{(3)}(\omega_\sigma)|\Ef_p|^2\Ef_s
\end{align}
The gain $g_s$ in \cite{Johnston:1968} is defined as the power exponential gain $e^{g_sz}$, so we look for solutions $\frac{dI_s}{dz}=g_sI_s$, where $I_s=\varepsilon_0 n_s c |\Ef_s|^2/2$ is the intensity. Since $\frac{dI_s}{dz}=\tfrac{1}{2}\varepsilon_0 n_s c 2{\rm Re}[\Ef_s^*\frac{d\Ef_s}{dz}]$ we get that
\begin{align}
    g_s&=-\frac{3\omega_s}{2n_sc}{\rm Im}[\chi^{(3)}(\omega_s)]|\Ef_p|^2
    \\&=
    -\frac{3\omega_s}{n_sn_p\varepsilon_0 c^2}{\rm Im}[\chi^{(3)}_{R}(\omega_s)]I_p
    \label{eq:GoI-chi3-1}
    \\&=
    -\frac{6\pi}{\lambda_sn_sn_p\varepsilon_0 c}{\rm Im}[\chi^{(3)}_{R}(\omega_s)]I_p
    \label{eq:GoI-chi3}
\end{align}
where we define the Kerr nonlinearity in frequency domain from Eq. (\ref{eq:chi3-el-Raman-frequency}) as $\chi^{(3)}(\omega)=\chi^{(3)}_E+\chi^{(3)}_R(\omega)$.
Eq. (\ref{eq:GoI-chi3-1}) is identical to \cite[Eq. (7.36)]{butcher:1990}. Since the Raman susceptibility is imaginary and negative at the Raman resonance frequencies, the Stokes wave experiences Raman gain, as expected. Also note that since the gain is given by the imaginary part of the nonlinear susceptibility, then the electronic part $\chi^{(3)}_E$ will not affect the result as it is real. We can now use this result to calculate the imaginary Raman peak values as
\begin{align}
    \chi_{R,\sigma}^{(3)}({\rm peak})&\equiv |{\rm Im}[\chi^{(3)}_{R}(\omega_s=\omega_p-\omega_\sigma)]|
\nonumber\\
\label{eq:chi3-GoI}    &=\frac{\lambda_sn_sn_p\varepsilon_0 c}{6\pi}\frac{g_s}{I_p}
\end{align}
Eq. (\ref{eq:chi3-GoI}) is a factor of 2 smaller than Eq. (10) in \cite{Phillips:2011}; it could be because they calculate the amplitude gain $\frac{d\Ef_s}{dz}=\alpha_s\Ef_s$ [this is at least the approach used in \cite{boyd:2007}, cf. Eq. (10.3.21) there -- but note that he uses a completely different definition of the electric fields and susceptibilities], and as is well known $\alpha_s=g_s/2$. Carrying out the calculations, we obtain the peak values \cite{Kaminow:1967,Kaminow:1969}
\begin{align}\label{eq:chi3-sigma-JK}
\chi_{R,\sigma}^{(3)}({\rm peak})=&\{5.2, 2.5, 0.61,4.8\}\times 10^4 ~\rm pm^2/V^2, \\
\nonumber &\text{(
using 30\% larger $\gamma_\sigma$)}
\end{align}
Assuming again a Lorentzian-based response for the Raman we can write it on the form
\begin{align}\label{eq:chi3R-JK}
    \chi_R^{(3)}(\Omega)&=\sum_{\sigma=1}^4 \chi_{R,\sigma}^{(3)}
    \frac{\omega_\sigma^2}{\omega_\sigma^2 +i 2\gamma_\sigma \Omega-\Omega^2}\\
    &=\chi_{R}^{(3)}\sum_{\sigma=1}^4 f_{R,\sigma}
    \frac{\omega_\sigma^2}{\omega_\sigma^2 +i 2\gamma_\sigma \Omega-\Omega^2}
\end{align}
where as usual $\chi_{R}^{(3)}=\sum_{\sigma}\chi_{R,\sigma}^{(3)}$ and we have used the requirement $-{\rm Im}[\chi_R^{(3)}(\omega_\sigma)]=\chi_{R,\sigma}^{(3)}({\rm peak})$ to define
\begin{align}\label{eq:chi3R-JK-tot}
\chi_{R,\sigma}^{(3)}({\rm peak})&=\frac{\chi_{R,\sigma}^{(3)}2\gamma_\sigma} {\omega_\sigma}+ \sum_{j\neq \sigma}    \frac{\omega_\sigma\omega_j^22\gamma_j\chi_{R,j}^{(3)}}
{(\omega_j^2-\omega_\sigma^2)^2 + (2\gamma_j)^2 \Omega_\sigma^2}\\
\label{eq:chi3R-JK-fRsigma}
f_{R,\sigma}&=\frac{\chi_{R,\sigma}^{(3)}}{\chi_{R}^{(3)}}
\end{align}
Note that knowing $\chi_{R,\sigma}^{(3)}({\rm peak})$ we can find the individual $\chi_{R,\sigma}^{(3)}$ by solving the system of equations for all $\sigma$ that Eq. (\ref{eq:chi3R-JK-tot}) defines. These definitions ensure that the peak values match the calculated ones $\chi_{R,\sigma}^{(3)}(\text{peak})$, and that everything is normalized as before. The calculations give \cite{Kaminow:1967,Kaminow:1969}
\begin{align}
    \label{eq:chi3-JK-tot-value}
    \chi_{R}^{(3)}&=0.62\times 10^4~\rm pm^2/V^2\\
    \label{eq:n2-JK-tot-value}
    n^I_{2,R}&=34.3\times 10^{-20}~\rm m^2/W\\
    \label{eq:fRsigma-JK-value}
    f_{R,\sigma}&=\{0.635,0.105,0.020,0.240\}
\end{align}
Note that these values are quite insensible to the linewidths used. 
The time constants are \cite{Kaminow:1967,Kaminow:1969}
\begin{align}
    \label{eq:tau1-JK-value}
    \tau_{1,\sigma}&=\{21.0, 19.3, 15.9, 8.3\}~\rm fs
\\
    \label{eq:tau2-JK-value}
    \tau_{2,\sigma}&=\{544, 1021, 1361, 544\}~\rm fs
\\    \nonumber &\text{(
using 30\% larger $\gamma_\sigma$)}
\end{align}
Finally we mention the ratio \cite{Kaminow:1967}
\begin{align}\label{eq:DC-ratio-KJ}
 \frac{\chi_{R}^{(3)}}{{\rm max}[\chi_{R,\sigma}^{(3)}({\rm peak})]}=&0.119
\\    \nonumber &\text{(
using 30\% larger $\gamma_\sigma$)}
\end{align}
This represents the ratio between the DC value and the peak of all the Raman lines (i.e. the value at zero frequency when the Raman spectrum is normalized to have its peak at unity).

Let us address the question about the measurements performed in the subsequent paper \cite{Johnston:1968,Kaminow:1969}, where the 256 and $637\icm$ modes at 300 K were reported to have linewidths of $2\gamma=23$ and $20\icm$, respectively, with the same reported scattering efficiencies.
Inserting the numbers of Table II of \cite{Johnston:1968} in Eq. (\ref{eq:GoI}) we get \cite{Johnston:1968,Kaminow:1969}
\begin{align}\label{eq:GoI-JK-1969}
    g_s/I_p=&\{12.5,12.9\}~{\rm cm/GW} 
\end{align}
at 256 and $637\icm$, respectively, in agreement with Table II of \cite{Johnston:1968} when correcting for the factor of 2 mistake \cite{Kaminow:1969}. Using the same approach as above the peak values of the two modes are calculated to be \cite{Johnston:1968,Kaminow:1969}
\begin{align}\label{eq:chi3-sigma-JK-1969}
    \chi_{R,\sigma}^{(3)}({\rm peak})=&\{4.4,4.6\}\times 10^4 ~\rm pm^2/V^2
\end{align}
These values agree quite nicely with their previous results, Eq. (\ref{eq:chi3-sigma-JK}), when all the corrections are applied. We can still use the value from Eq. (\ref{eq:chi3-JK-tot-value}), as calculated from their previous measurements, as the DC value, because only the linewidths separate the measurements, and as mentioned above they do not influence the DC value. Assuming the stronger of these two lines to be the peak of all the Raman lines, the ratio of the DC value and the peak of all the Raman lines is now \cite{Johnston:1968,Kaminow:1969}
\begin{align}\label{eq:DC-ratio-JK-1969}
 \chi_{R}^{(3)}/{\rm max}[\chi_{R,\sigma}^{(3)}({\rm peak})]=0.133
\end{align}
again close to the previous result Eq. (\ref{eq:DC-ratio-KJ}).

Johnston \cite{Johnston:1968a} measured the gain at $\lambda=1.064\mic$ of the $\nu_\sigma=\{256,637\}\icm$ lines as
\begin{align}\label{eq:GoI-Johnston}
    g_s/I_p=\{5.75,6.0\}~\rm cm/GW,
    \\\nonumber \text{(
    correcting for a factor 2)}
\end{align}
Calculating the peak values of the Raman susceptibility and assuming the same linewidths as in \cite{Johnston:1968} (although it is not specifically defined) we get \cite{Johnston:1968a}
\begin{align}\label{eq:chi3Rsigma-Johnston}
    \chi_{R,\sigma}^{(3)}({\rm peak})=\{4.1,4.5\}\times 10^4 ~\rm pm^2/V^2     \\\nonumber \text{(
    correcting for a factor 2)}
\end{align}

\section{Other Raman models and Raman gain measurements}
\label{sec:other}

A measurement of the two main Raman gain peaks is reported by Boyd \cite[p. 480]{boyd:2007}, where the $\nu_\sigma=\{256,637\}\icm$ lines are reported to have linewidths $\Delta\nu_\sigma=\{23,20\}\icm$, and peak gain
\begin{align}\label{eq:GoI-Boyd}
    g_s/I_p=\{8.9,9.4\}~\rm cm/GW, 
\end{align}
measured at 694 nm. Calculating the peak values of the Raman susceptibility and assuming that $\Delta\nu_\sigma=\Gamma_\sigma$ (although it is not specifically defined) we get
\begin{align}\label{eq:chi3Rsigma-Boyd}
    \chi_{R,\sigma}^{(3)}({\rm peak})=\{4.2,4.6\}\times 10^4 ~\rm pm^2/V^2 
\end{align}

Finally, in Chunaev et al. \cite{Chunaev2006} measured the gain at $\lambda=1.047\mic$ of the $256\icm$ mode 
\begin{align}\label{eq:GoI-Chunaev}
    g_s/I_p=5.1~\rm cm/GW 
\end{align}
The decay time of the mode was found as $\tau_2=0.38$ ps, which gives a linewidth of $\Gamma=28\icm$.
Calculating the peak values of the Raman susceptibility we get
\begin{align}\label{eq:chi3Rsigma-Chuanev}
    \chi_{R,\sigma}^{(3)}({\rm peak})=3.6\times 10^4 ~\rm pm^2/V^2
\end{align}

\begin{figure}[tb]
\includegraphics[width=8.5cm]{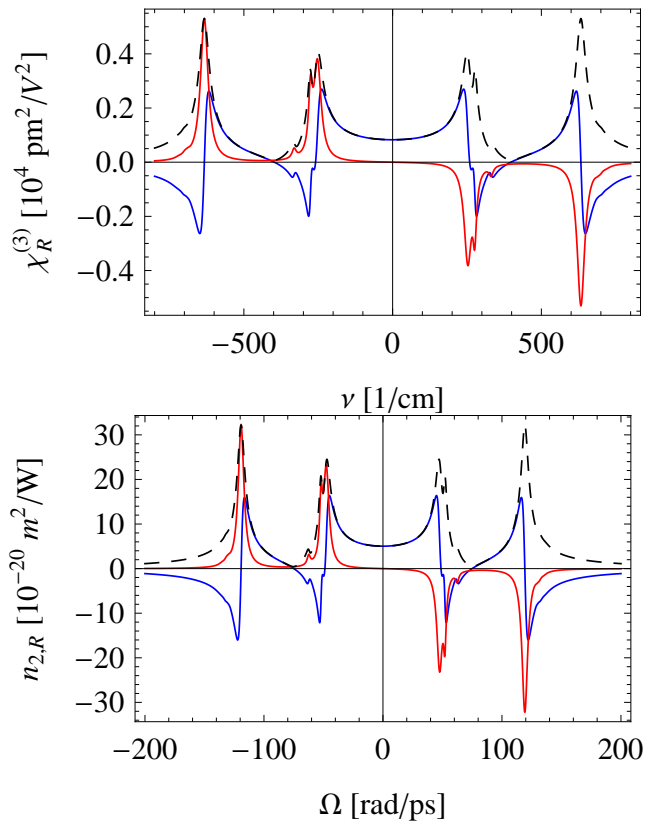}
\caption{\label{fig:Raman-Phillips} Data from Phillips et al. \cite{Phillips:2011}. The Raman spectrum is scaled to give the peak value of ${\rm max}[\chi_{R,\sigma}^{(3)}({\rm peak})]=5.3\times 10^{-21}~\rm m^2/V^2$ of the largest frequency mode, applicable for $\lambda=1.043\mic$.
    }
\end{figure}

We include for completeness the Raman spectrum of Phillips et al. \cite{Phillips:2011} in Fig. \ref{fig:Raman-Phillips}. They found 7 modes, the first four are the standard ones plus 3 minor modes. They reported a DC ratio
\begin{align}
\label{eq:DC-ratio-Phillips}
 \frac{\chi_{R}^{(3)}}{{\rm max}[\chi_{R,\sigma}^{(3)}({\rm peak})]}&=0.17
\\
\label{eq:DC-ratio-Phillips-calculated}
&=0.156
\end{align}
where the latter value was found from calculating the ratio from their Table 1. The 3 extra modes they include do not contribute much to this ratio. The chose to use ${\rm max}[\chi_{R,\sigma}^{(3)}({\rm peak})] =5.3\times 10^{-21}~\rm m^2/V^2$ for $\lambda=1.043\mic$. This  was determined based on data from numerical simulations. We then get using the DC ratio 0.17
\begin{align}
    \label{eq:chi3-Phillips-tot-value}
    \chi_{R}^{(3)}&=0.090\times 10^4~\rm pm^2/V^2\\
    \label{eq:n2-Phillips-tot-value}
    n^I_{2,R}&=5.5\times 10^{-20}~\rm m^2/W
\end{align}
When comparing to the previous measurements these values are very low. Note also how strong the large-frequency mode is relative to the main low-frequency mode. Finally, note that the spectral form they use $a_j/(f_j^2-f^2+2i\gamma_jf)$ is not normalized.

\begin{figure}[tb]
\includegraphics[width=8.5cm]{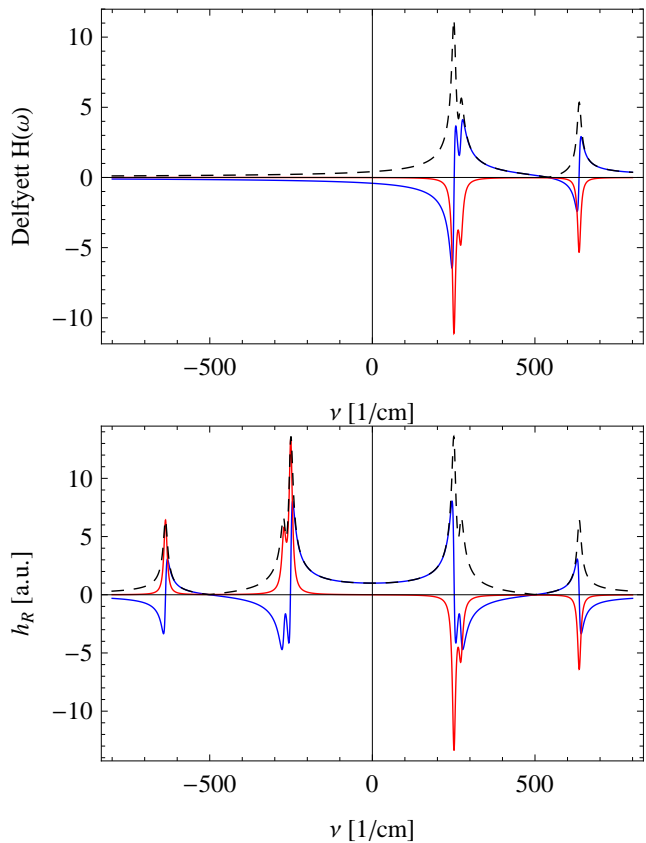}
\caption{\label{fig:Raman-Delfyett} Raman spectrum from Delfyett et al. \cite{Delfyett:1989} measured for $\lambda=0.532\mic$, and using $a_i$ as the oscillator strengths. Top: their model, Eq. (\ref{eq:chi3R-Delfyett}). Bottom: equivalent model using the symmetric Lorentzian Eq. (\ref{eq:hR-omega}). Notice that the real part of the top plot is negative in the model they use.
    }
\end{figure}

Delfyett \cite{Delfyett:1989} is again another story, because they fit to what they claim to be a Lorentzian and get the Raman contribution to the total nonlinearity of
\begin{align}\label{eq:chi3R-Delfyett}
    H(\omega)=\sum_{\sigma=1}^3 \frac{a_\sigma}{\omega-\omega_\sigma+j\gamma_\sigma}
\end{align}
This form is problematic because (a) it is not (conjugate) symmetric towards negative frequencies (b) the real part is negative at zero frequency. We remark that one can reduce the classical Lorentzian form Eq. (\ref{eq:hR-omega}) to this form close to a resonance. Let us transform their form into the standard form, Eq. (\ref{eq:hR-omega}) by normalizing it. Thus, with the parameters
\begin{align}\label{eq:delfyett-fR}
    f_{R,\sigma}&=\frac{a_\sigma}{\omega_\sigma\sum_\sigma a_\sigma/\omega_\sigma} \\
\label{eq:delfyett-fR-val}
&=\{0.61,0.27,0.12\}
\end{align}
and obviously $2\gamma_\sigma=\Gamma_\sigma$ we can use the standard form. We have in Fig. \ref{fig:Raman-Delfyett} used as $a_\sigma$ and $\gamma_\sigma$ the parameters $a_i$ and $\Gamma_i$ in Table I of \cite{Delfyett:1989}. We also find the DC ratio to peak as
\begin{align}
\label{eq:DC-ratio-Delfyett}
 \frac{\chi_{R}^{(3)}}{{\rm max}[\chi_{R,\sigma}^{(3)}({\rm peak})]}&=0.075
\end{align}

\begin{figure}[tb]
\includegraphics[width=8.5cm]{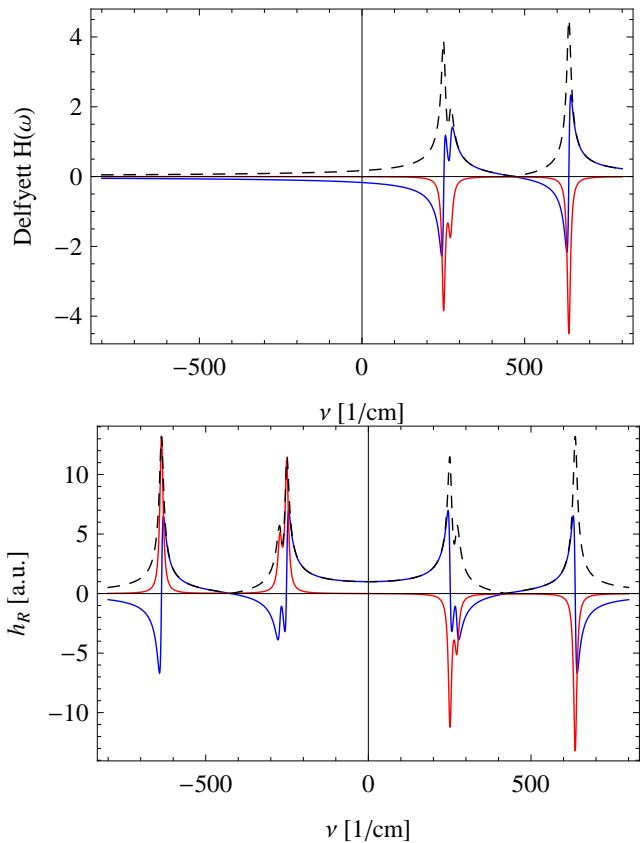}
\caption{\label{fig:Raman-Delfyett1} As Fig. \ref{fig:Raman-Delfyett} but using $a/\chi_{\rm nr}^{(3)}$ as the oscillator strengths.
    }
\end{figure}


\section{Frequency dependence of the electronic and Raman susceptibilities}
\label{sec:freq}

Now we have $\chi_{R}^{(3)}$, but in order to estimate the Raman fraction $f_R$ we need to know the total Kerr nonlinearity (or the electronic Kerr nonlinearity), and $f_R=\chi_{R}^{(3)}/\chi_{\rm tot}^{(3)}$. At this point it is relevant to discuss the frequency dependence of the nonlinearities. We know that the nonlinear susceptibilities change with frequency. Miller \cite{miller:1964} first pointed out that $\delta\equiv \chi^{(2)}(\omega_1,\omega_2,\omega_3) /[\chi^{(1)}(\omega_1)\chi^{(1)}(\omega_2)\chi^{(1)}(\omega_3)]$ is roughly constant, hence it is known as Miller's $\delta$. One can therefore given the value at one wavelength apply Miller's scaling to estimate the nonlinear coefficient at a new wavelength. Such a relation also holds for the higher-order nonlinearities. It can be derived from treating the atom and the electron using the Lorentz model of a classical anharmonic oscillator, and then including a nonlinear force of the electron (see e.g. \cite{boyd:2007}). Therefore one can only use this approach to scale the electronic nonlinear susceptibility.

\begin{figure}[tb]
\includegraphics[width=8.5cm]{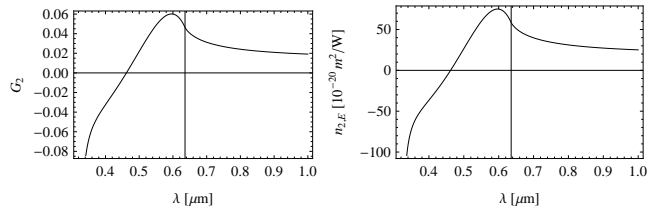}
\caption{\label{fig:G2} Left: The degenerate $G_2$ parameter \cite[p. 279]{Sheik-Bahae:1998} of LN vs. wavelength, calculated using $E_g=3.9$ eV \cite{desalvo:1996}. To a good approximation the frequency dependence of the electronic nonlinearity $n^I_{2,E}\propto G_2$. The vertical line at 636 nm marks the wavelength of a photon carrying half the bandgap energy. Right: calculated $n^I_{2,E}$ from Eq. (\ref{eq:n2-TBM}) using the $G_2$ in the top graph.
    }
\end{figure}

A more accurate model is the two-band model, which is \cite[Eq. (55)]{Sheik-Bahae:1998}
\begin{align}\label{eq:n2-TBM}
    n^I_{2,E}(\omega)\text{(SI)}=K'\frac{\sqrt{E_p}}{n^2(\omega) E_g^4}G_2(\hbar \omega/E_g)
\end{align}
where $K'$ is a material constant, fitted to $K'=7.3\times 10^{-9}~{\rm eV^{3.5}m^2/W}$ in wide-gap dielectrics \cite{desalvo:1996}, and $E_p=21$ eV is the Kane energy, which is constant for most materials. With these two constants one can a priori estimate the electronic nonlinearity, in particular the enhancement close to the two-photon absorption regime. The curves are plotted in Fig. \ref{fig:G2}.


The question is now: does the DC Raman nonlinearity $\chi^{(3)}_R$ scale with wavelength? And if yes, how does it scale?

In order to answer this, consider a classical description where Raman scattering can be modeled as a simple harmonic oscillator. Consider the standard case (no Raman): the electric field induces a change in the molecular dipole moment $p=\varepsilon_0 \alpha E$ by shifting the electrons away from the nucleus (consistent with before we use tilde to denote the choice where $\varepsilon_0$ is kept apart from the polarizability, as we below have another case where this is not done). The linear susceptibility is given by $\chi^{(1)}=N \alpha(\omega)$, where $N$ is the atom number density, and the polarizability has a Lorentz form
\begin{align}\label{eq:pol-noRaman}
     \alpha(\omega)=\frac{1}{\varepsilon_0}\frac{e^2/m} {\omega_\sigma^2-\omega^2-2i\gamma_\sigma\omega}
\end{align}
where $e$ is the electron charge, $m$ the electron mass, $\omega_\sigma$ the dipole resonance frequency and $\gamma_\sigma$ the dipole damping rate. This approach assumes that $\alpha=\alpha_0$ is constant in time, and therefore the electrical-field shifted electron cloud is in equilibrium at its new shifted position away from the nucleus. The Raman effect occurs when the polarizability has a temporal dependence around the equilibrium value \cite{boyd:2007} $\alpha_0$ as $ \alpha (t)= \alpha_0+ q(t)(\partial \alpha/\partial q)_0$, where $(\partial \alpha/\partial q)_0$ is the (first) hyperpolarizability \cite[Ch. 15]{handbook-IV:2010}, also called the differential polarizability \cite[p. 199]{butcher:1990}, and where the subscript '0' indicates that it is a static value at the particular applied optical frequency. The Stokes field turns out to have the following nonlinear susceptibility \cite[Eq. (10.3.19a)]{boyd:2007}
\begin{align}\label{eq:chi3R-stokes-Boyd}
    \chi_R^{(3)}(\omega_s)=\frac{\varepsilon_0(N/6m)(\partial \alpha/\partial q)_0^2} {\omega_\sigma^2-(\omega_p-\omega_s)^2+2i\gamma_\sigma(\omega_p-\omega_s)}
\end{align}
Thus, at zero frequency we have $\chi_R^{(3)}(\omega_s=\omega_p)= \varepsilon_0 N(\partial \alpha/\partial q)_0^2/(6m\omega_\sigma^2)$. The only term that could have a frequency dependence is the $(\partial \alpha/\partial q)_0$ term, but in the classical model it is a purely phenomenological quantity so we cannot from this approach say anything. Certainly $\alpha$ depends on frequency, this is stated in Eq. (\ref{eq:pol-noRaman}). But is the hyperpolarizability $(\partial \alpha/\partial q)_0$ dispersive? According to \cite[p. 199]{butcher:1990} the Raman susceptibility close to a resonance can according to a classical derivation be stated as
\begin{align}\label{eq:ch3R-Butcher-classical}
       \chi_R^{(3)}(\omega_s)=\frac{N|(\partial \bar\alpha/\partial X)_0|^2} {6\hbar\varepsilon_0(\omega_\sigma-\omega_p+\omega_s+i\gamma_\sigma)}
\end{align}
where $(\partial \bar\alpha/\partial X)_0$ is the differential polarizability, and as the notation indicates it is defined in a different way than $(\partial \alpha/\partial q)_0$, see more below. In the same limit Eq. (\ref{eq:chi3R-stokes-Boyd}) becomes \cite[Eq. (10.3.19b)]{boyd:2007}
\begin{align}\label{eq:chi3R-stokes-Boyd-app}
    \chi_R^{(3)}(\omega_s)=\frac{\varepsilon_0N(\partial \alpha/\partial q)_0^2} {12m\omega_\sigma(\omega_\sigma-\omega_p+\omega_s+i\gamma_\sigma)}
\end{align}
Care must be taken in comparing these two results: in \cite{butcher:1990} the dipole moment $p$ induced by an electric field $E$ defines the polarizability as $p=\bar\alpha E$. Instead in \cite{boyd:2007} it is defined as $p=\varepsilon_0\alpha E$; this ensures that $\alpha$ has units $[\rm m^3]$ and that $\chi^{(1)}=N \alpha(\omega)$ holds. We of course have simply that $ \bar\alpha=\varepsilon_0\alpha$, and that $[\bar\alpha]=\rm Fm^2$. More importantly, in \cite{butcher:1990} $X$ is a dimensionless coordinate, so $(\partial \bar\alpha/\partial X)_0$ has the same units as $\bar\alpha$. This means that the (reduced) electron mass $m$ used above is replaced by $\hbar/2\omega_\sigma$, stated in \cite{butcher:1990} as a reduced mass but really is a reduced mass normalized to a characteristic length suitable for normalization of $X$. In essence if one makes the substitution $\sqrt{\hbar/2m\omega_\sigma}X=q$ the two results become identical. In \cite{butcher:1990} the classical derivation leads to the following identity $(\partial \bar\alpha/\partial X)_0=\bar\alpha_{fg}^{\rm tr}(-\omega_s;\omega_p)$, linking the result to a quantum mechanical derivation where the first order transition hyperpolarizability, describing the generation of a Stokes wave, was introduced as \cite[Eq. (4.115)]{butcher:1990}
\begin{align}\label{eq:dalpha-dq-QM}
    \bar\alpha_{fg}^{\rm tr}(-\omega_s;\omega_p)=\frac{e^2}{\hbar}\sum_i\left[
    \frac{\mathbf{e}_s\cdot \mathbf{r}_{fi}\mathbf{e}_p\cdot \mathbf{r}_{gi}} {\Omega_{ig}-\omega_p}+
    \frac{\mathbf{e}_p\cdot \mathbf{r}_{fi}\mathbf{e}_s\cdot \mathbf{r}_{gi}} {\Omega_{ig}+\omega_s}
    \right]
\end{align}
where the subscript $fg$ symbolizes the the transition from a molecular ground state $g$ to a excited final state $f$ with transition frequency $\Omega_{fg}$, and the sum over the index $i$ refers to all the final states as observed from the molecular ground state $g$. We have here used the superscript 'tr' to emphasize that this transition hyperpolarizability is different than the molecular polarizabilities $\bar\alpha$ (or $\alpha$). They are defined in similar ways but with different scopes: $\bar\alpha$ is the first order expansion coefficient in the dipole moment when Taylor expanded around the applied electric field (e.g., $p=p_0+\bar\alpha E$) \cite[Eq. (4.83)]{butcher:1990}, while $\bar\alpha^{\rm tr}$ is the first order expansion coefficient in the polarizability when Taylor expanded around the vibrational position coordinate ($q$ or $X$) from its equilibrium value (e.g., $\bar\alpha=\bar\alpha_0+\bar\alpha^{\rm tr} X$) \cite[Eq. (6.121)]{butcher:1990}. The electric field is defined as $\mathbf{E}_j=\mathbf{e}_jE_j$, so $\mathbf{e}_j$ is the dimensionless unit vector. The vibrational coordinates of the Raman modes are given by $\mathbf{r}_j$, and $e$ is the charge of the electron. It is important to note that $[\bar\alpha_{fg}^{\rm tr}(-\omega_s;\omega_p)]=\rm Cm^2/V=Fm^2$, i.e. the same as $\bar\alpha$ and $(\partial \bar\alpha/\partial X)_0$. We can now compare the two cases and get
\begin{align}\label{eq:differential-pol-correspondence}
    |\left(\partial \bar\alpha/\partial q\right)_0|^2 &=\frac{2m\omega_\sigma}{\hbar}|\bar\alpha_{fg}^{\rm tr}(-\omega_s;\omega_p)|^2
\end{align}
and obviously $\left(\partial \bar\alpha/\partial q\right)_0=\varepsilon_0\left(\partial \alpha/\partial q\right)_0$. This shows that we can expect the differential polarizability to experience resonant enhancement when the pump photon $\omega_p$ or the Stokes photon $\omega_s$ comes close to a molecular resonance frequency. [Eq. (\ref{eq:dalpha-dq-QM}) is derived assuming interaction far from such a resonance, and thereby the damping terms have been removed from the denominators in the square brackets in Eq. (\ref{eq:dalpha-dq-QM}).] The numerators should actually express quantum mechanical expectation averages over the ground and final stage wave functions and the electric dipole moment operator. Whether these quantities, which could be calculated in the simplest case using e.g. variation-perturbation theory, vary with pump frequency is an open question.

We can now try to express the peak gain when $\omega_p-\omega_s=\omega_\sigma$ using the polarization derivative. The peak Raman susceptibility Eq. (\ref{eq:chi3R-stokes-Boyd-app}) becomes at this frequency
\begin{align}\label{eq:chi3R-stokes-Boyd-app-peak}
    \chi_R^{(3)}(\text{peak})=-i\frac{\varepsilon_0N(\partial \alpha/\partial q)_0^2} {12m\omega_\sigma\gamma_\sigma}
\end{align}
We can now use Eq. (\ref{eq:chi3-GoI}) to get the peak Raman gain coefficient as
\begin{align}\label{eq:gain-peak-alpha-1}
    \frac{g_s}{I_p}=\frac{\omega_s N (\partial \alpha/\partial q)_0^2} {4 m \omega_\sigma \gamma_\sigma n_sn_pc^2}
\end{align}
If we do the same by inserting Eq. (\ref{eq:scattering-cross-JK}) into Eq. (\ref{eq:GoI}) then we get the same result, provided that we substitute $(\partial \bar\alpha/\partial q)_0=\varepsilon_0(\partial \alpha/\partial q)_0$.

Hellwarth used the Born-Oppenheimer approximation to derive the nuclear (vibrational) contribution to $\chi^{(3)}$ and argued that it must scale as $\chi^{(3)}_R\propto\chi^{(1)}(\omega_p)\chi^{(1)}(\omega_s)$ \cite[Sec. 13.2.1]{hellwarth:1977}, which then implies
\begin{align}\label{eq:chi3-scaling-Hellwarth}
    \chi^{(3)}_R\propto(n_p^2-1)(n_s^2-1)\simeq (n_p^2-1)^2
\end{align}
which means that the Raman \textit{gain} scales as ${\rm Im}[\chi_R^{(3)}]/(\lambda_pn_pn_s)$, c.f. Eq. (\ref{eq:GoI-chi3}), or
\begin{align}\label{eq:gain-scaling-Hellwarth}
    g_s/I_p\propto(n_p^2-1)(n_s^2-1)/(\lambda_sn_sn_p)
\end{align}
This presumably is what leads to the simple scaling-factor of the Raman gain $(n^2-1)^2/n^2$ reported several places \cite{stolen:2000,Rivero2005,Stegeman2006}; what Stolen \cite{stolen:2000} refers to when quoting the $(n^2-1)^2/n^2$ scaling is the gain scaling besides the inverse scaling with the wavelength and he then presumably simplifies $n_p\simeq n_s$. However, it also seems a sort of conclusion from \cite{Rivero2005,Stegeman2006} that this scaling alone does not explain the measured increase in the "normalized" Raman gain (normalized to scale out the $\lambda_p^{-1}$ dependence) when the pump frequency is close to an electronic transition (i.e. the UV loss edge), see e.g. \cite[Fig. 4]{Rivero2005}. The increase was instead explained as a resonance enhancement \cite{Rivero2005,Stegeman2006}.

How can we model this resonance enhancement of the Raman susceptibility? Eq. (\ref{eq:dalpha-dq-QM}) shows a resonance enhancement, but this is connected to a proximity to the molecular transition frequencies, and not electronic transition frequencies. Levenson and Bloembergen \cite{Levenson:1974} expressed the Raman terms with the local-field corrected polarizability $\bar\alpha^R$, which corresponds to the first-order transition hyperpolarizability as defined in Eq. (\ref{eq:dalpha-dq-QM}); this can readily be checked since \cite[Eq. (8)]{Levenson:1974} corresponds to Eq. (\ref{eq:differential-pol-correspondence}) when neglecting what presumably is the local-field correction term $[(n^2+1)/3]^2$ (although the standard field-correction factor is $[(n^2+2)/3]^2$, see \cite[Eq. (4)]{kato:1971}). Moving on, they introduced a phenomenological scaling as \cite[Eq. (18)]{Levenson:1974} $\bar\alpha^R\propto 1/(E_--\hbar\omega_p)$, where $E_-$ is the is the energy of the direct band gap. They do mention that the resonance energy is chosen rather arbitrarily, and could also be chosen to coincide with that of the Sellmeier equation. For LN we have $E_g=3.9$ eV, corresponding to $\lambda=0.32\mic$. The UV pole of the Sellmeier equation for $n_e$ instead lies around $\lambda=0.20-0.21\mic$. Kato and Takuma \cite{kato:1971} relied on a different quantum mechanical derivation of the Raman response by Peticolas et al. \cite{peticolas:1970}, and they arrived at a resonance enhancement of what corresponds to the transition hyperpolarizability as $R\propto [(\omega_0^2-\omega_p^2)(\omega_0^2-\omega_s^2)]^{-1}$, where $\omega_0$ is the energy difference between a ground and an excited electronic state without coupling to molecular vibration \cite[Eq. (3)]{kato:1971}. The total Raman gain was then given as $\propto |R|^2$. This suggests an enhancement of the Raman gain that scales as $[(\omega_0^2-\omega_p^2)(\omega_0^2-\omega_s^2)]^{-2}$, assuming a symmetric excitation of the vibration (I assume this means stimulated Raman scattering, i.e. that the two pump photons excite the same Raman mode, and thus the degenerate case of \cite[Fig.4.5(b)]{butcher:1990}). Other authors have without providing any reasoning chosen to model the scaling of the Raman gain coefficient as \cite{Bischel1986,Brasseur1998,lisinetskii:2005}
\begin{align}\label{eq:gS-scaling-empirical}
    \frac{g_s}{I_p}=D\frac{\nu_s}{(\nu_0^2-\nu_p^2)^2}
\end{align}
where $\nu_0$ is the resonance frequency. Similar results were also found in \cite[Eq. (2)]{albrecht:1971}. In fact, there they find using the Born-Oppenheimer approximation that the transition hyperpolarizability $\bar\alpha^{\rm tr}=A+B$, where the parameters $A\propto(\nu_e^2+\nu_p^2)/(\nu_e^2-\nu_p^2)^2$, corresponding to a diagonal "symmetric" excitation, and $B\propto(\nu_e\nu_s+\nu_p^2)/(\nu_e^2-\nu_p^2)(\nu_s^2-\nu_p^2)$, corresponding to an off-diagonal "asymmetric" excitation, and $\nu_e$ and $\nu_s$ are virtual electronic states (i.e. energy differences between ground and excited electronic states). The total Raman cross section then scales as $\nu_s^4|\bar\alpha^{\rm tr}|^2$ so they introduce two dimensionless frequency factors
\begin{align}
\label{eq:Raman-scaling-Albrecht-FA}
    F_A&=\nu_s^2(\nu_e^2+\nu_p^2)/(\nu_e^2-\nu_p^2)^2 \\
\label{eq:Raman-scaling-Albrecht-FB}
    F_B&=2\nu_s^2(\nu_e\nu_s+\nu_p^2)/(\nu_e^2-\nu_p^2)(\nu_s^2-\nu_p^2)
\end{align}
Depending on the nature of the experiment, the Raman scattering cross section can now scale as $F_A^2$, corresponding to a symmetric excitation, as $F_B^2$, corresponding to an asymmetric excitation, or a mixture, $(F_A+F_B)^2$.

\begin{figure}[tb]
\includegraphics[height=3cm]{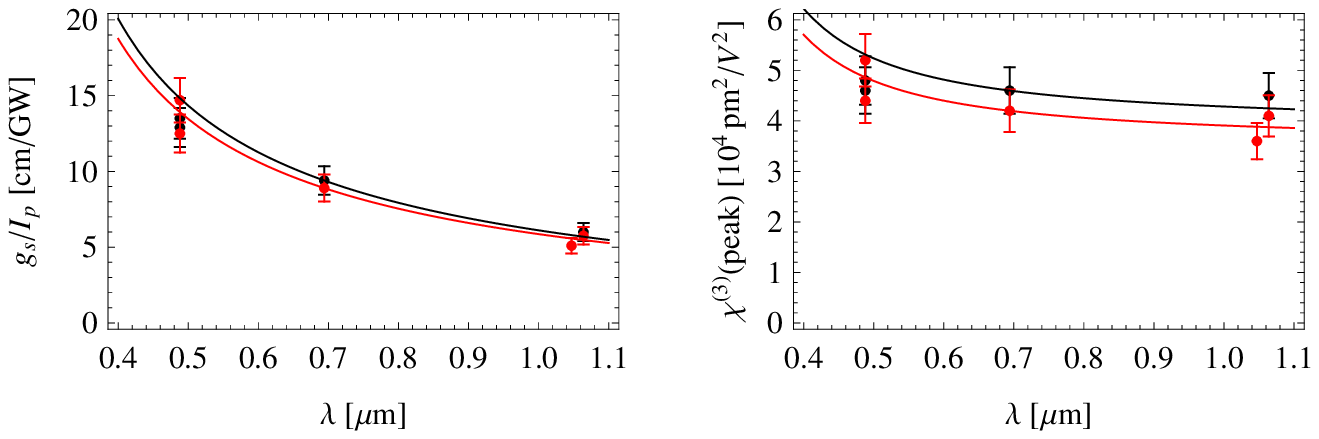}
\includegraphics[height=4cm]{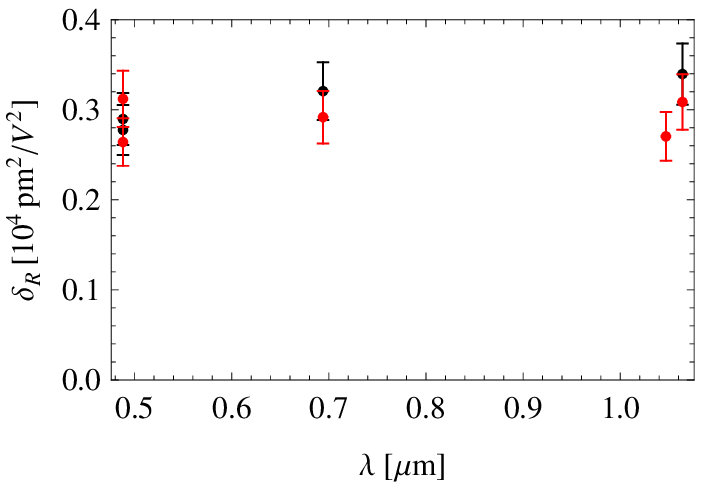}
\caption{\label{fig:gain} Top left: Raman Stokes gain $g_s/I_p$ as measured at various wavelengths. The red (black) dots are the $256 \icm$ ($637\icm$) mode, and the curves follow the scaling suggested by Eq. (\ref{eq:gain-scaling-Hellwarth}), and is chosen to be centered around the measurement at 694 nm.  Top right: the same measurements expressed through the peak Raman susceptibility $\chi_R^{(3)}(\text{peak})$. The curves now follow the scaling suggested by Eq. (\ref{eq:chi3-scaling-Hellwarth}). Any resonance scaling was not found to match the data. Bottom: suggested frequency independent parameter $\delta_R$, cf. Eq. (\ref{eq:deltaR}) vs. wavelength. The error bars are indicating an estimated 10\% standard deviation. 
    }
\end{figure}

In Fig. \ref{fig:gain} we plot the gain data vs. wavelength, and we found surprisingly that the resonant enhancement scaling laws did not fit very well with the data. Concerning the Raman gain, left plot, we found that the simple scaling found by Hellwarth, Eq. (\ref{eq:gain-scaling-Hellwarth}), seems to fit. However, a large part of the wavelength dependence is attributed to the $1/\lambda_s$ dependence, so a more accurate way of understanding the frequency dependence is to plot the peak susceptibility, which we have done in the right plot. The red data, i.e. the low-frequency Raman mode, seems to scale like suggested, i.e. as Eq. (\ref{eq:chi3-scaling-Hellwarth}), while the black data from the high-frequency mode seems not to depend on wavelength at all. In order to remove any wavelength dependence one could suggest a sort of Raman-version of the Miller's delta as
\begin{align}\label{eq:deltaR}
    \delta_R\equiv \frac{\chi^{(3)}_R}{\chi^{(1)}(\omega_p)\chi^{(1)}(\omega_s)}
\end{align}
which then should remain constant with wavelength. We have plotted these data in the bottom figure, and obviously the scaling seems reasonable for the low-frequency mode, while the high-frequency mode seems to increase with wavelength. It should here also be noted that for some strange reason there is no agreement in the literature whether the low-frequency mode should be stronger than the high-frequency mode \cite{barker:1967,Kaminow:1967,Schiek:2005,Lengyel:2007,Ridah:1997,Schwarz:1997}, or vice versa \cite{Johnston:1968,Johnston:1968a,Phillips:2011,Delfyett:1989}; note in this connection that above room temperature the high-frequency mode becomes stronger \cite{Schwarz:1997}. This uncertainty in determining the mode strengths could influence the analysis of the data. Again we stress that we did not find any resonance enhancement for the data, which could be because the possible resonances lie too far into the UV to be observed.

\section{Measurements of the Kerr nonlinearity}
\label{sec:zscan}


In DeSalvo et al. \cite{desalvo:1996} they used a $Z$-scan technique to measure $n_2^E=48\pm 7\times 10^{-14}$ esu ($n^I_2=9.1\pm1.3\times 10^{-20}~{\rm m^2/W}$) at $\lambda_1=1.064\mic$ in an $X$-cut congruent LN crystal (undoped) with the beam polarized along $Z$, which means they measured the $\chi_{ZZZZ}^{(3)}$ tensor component. They assumed that no cascading contributed to this value, since the phase mismatch is large. We calculate using the Sellmeier equation for congruent LN of \cite{Jundt:1997} that $n_e(\omega_1)=2.1558$ and $n_e(2\omega_1)=2.23421$, so $\Delta k=926\imm$. We now use the well-known quadratic cascading Kerr-like nonlinear refractive index \cite{desalvo:1992}
\begin{align}\label{eq:n2DeSalvo}
    n^I_{2,\rm casc}&=-\frac{4\pi d_{\rm eff}^2}{\lambda_p\varepsilon_0 c n_1^2n_2 \Delta k}
\end{align}
where $d_{\rm eff}$ is the effective quadratic nonlinearity in units $[\rm m^2/V^2]$ and $\Delta k=k_2-2k_1$, $k_j=\omega_j n_j/c$, and $n_j$, $j=1,2$ are the linear refractive indices of the pump ($j=1$) and the second harmonic ($j=2$).
With $d_{\rm eff}=d_{33}=25.2$ pm/V \cite{Shoji:1997} we get from Eq. (\ref{eq:n2DeSalvo}) that $n^I_{2,\rm casc}=-29.4\pm2.9\times 10^{-20}~{\rm m^2/W}$; the uncertainty stems from a $\pm5\%$ accuracy in $d_{33}$ \cite[p. 2272]{Shoji:1997}. Thus, cascading contributes quite strongly to the measured nonlinear refractive index. A 30 ps pulse duration was used, so they measured the electronic plus the DC value of the Raman term, see also Eq. (\ref{eq:chi3-zscan}). If we subtract the estimated cascading value, the actual total Kerr nonlinear index would be
\begin{align}\label{eq:n2-LN-desalvo-corrected}
    n^I_{2,\rm tot}&=38.6\pm 3.2\times 10^{-20}~{\rm m^2/W} 
\end{align}
where the error has been calculated using propagation of uncertainty $\sigma_{\rm tot}^2=\sigma_{\rm Z-scan}^2+\sigma_{\rm casc}^2$; the relative error is 8.4\%.

DeSalvo et al. also performed the $Z$-scan measurement at $\lambda=532$ nm and measured $n_2^E=440\pm 70\times 10^{-14}$ esu ($n^I_{2}=83\pm13\times 10^{-20}~\rm m^2/W$). The cascading contribution at this wavelength is small, but not insignificant. To estimate it we calculate using the Sellmeier equation for congruent LN of \cite{Jundt:1997} that $n_e(\omega_1)=2.23421$ and $n_e(2\omega_1)=2.28590$, so $\Delta k=14,759\imm$. We then use Miller's scaling to translate $d_{\rm eff}=d_{33}=-25.7$ pm/V measured at $\lambda_1=0.852\mic$ \cite{Shoji:1997} to 532 nm and get $d_{\rm eff}=-49.6$ pm/V. Thus using Eq. (\ref{eq:n2DeSalvo}) we have $n^I_{2,\rm casc}=-10.4\pm3.1\times 10^{-20}~{\rm m^2/W}$. The uncertainty of this number lies mostly in the $d_{\rm eff}$ value (we estimate at least 10\% error from Miller's scaling on top of the 5\% accuracy of the measurement itself), because although the Sellmeier linear refractive index at 266 nm is not guaranteed to be that accurate, we checked that UV measurements \cite{palik:1998} support the calculated Sellmeier value. The total Kerr nonlinearity is then when correcting for cascading
\begin{align}\label{eq:n2-LN-desalvo-corrected-532}
    n^I_{2,\rm tot}&=93\pm14\times 10^{-20}~{\rm m^2/W}
\end{align}
Such a high value seems surprising, but consider that the wavelength corresponds to a photon energy of 2.33 eV, or 60\% of the bandgap value in LN. In this regime a strong 2PA enhancement of the electronic nonlinearity is predicted by the two-band model (the predicted peak lies just beyond half the bandgap energy \cite[Fig. 1]{sheik-bahae:1990}). The relative error on this value is 15\%, mainly determined by the error of the $Z$-scan measurement.

\begin{figure}[tb]
\includegraphics[width=8.5cm]{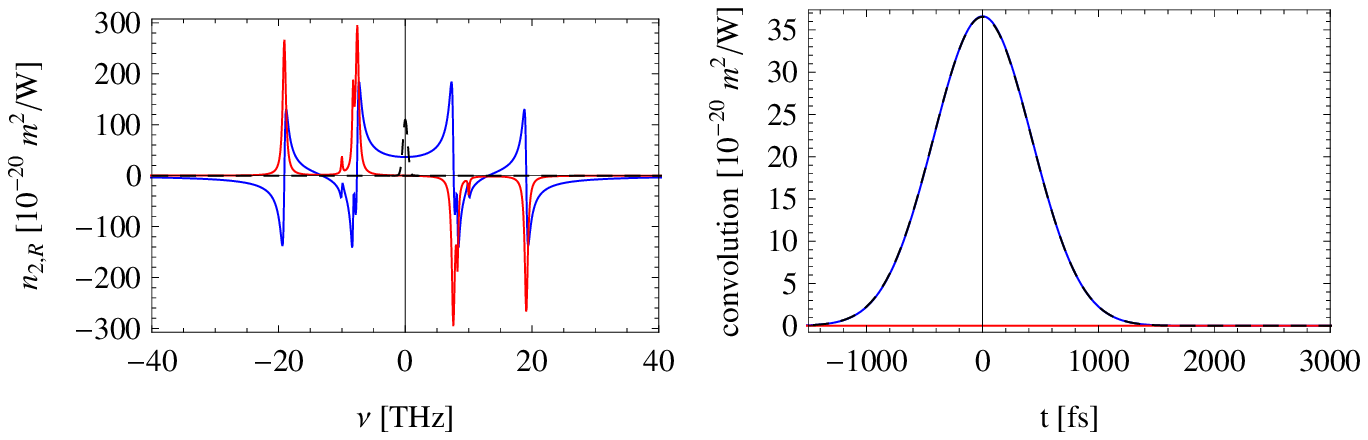}
\includegraphics[width=8.5cm]{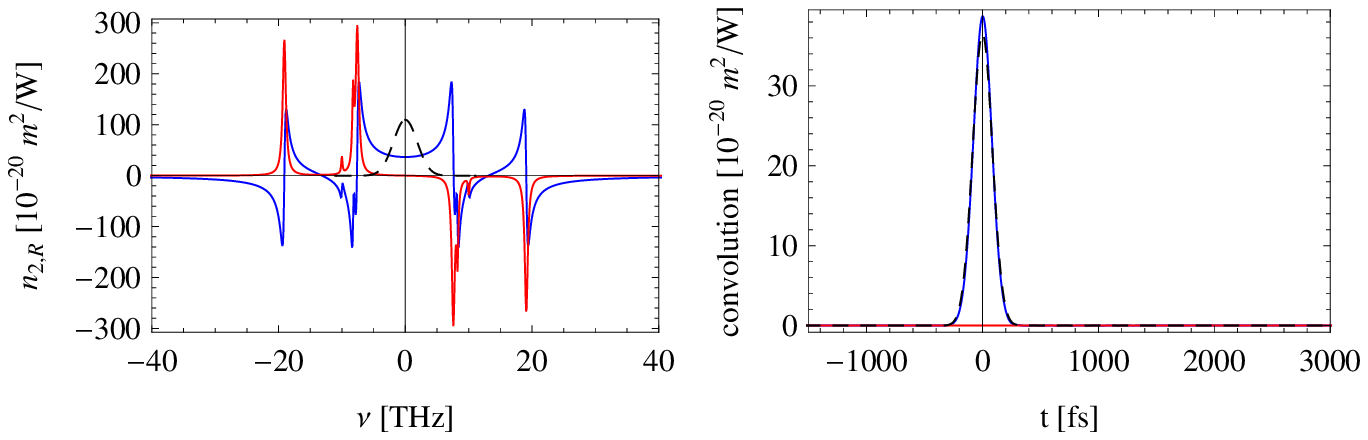}
\includegraphics[width=8.5cm]{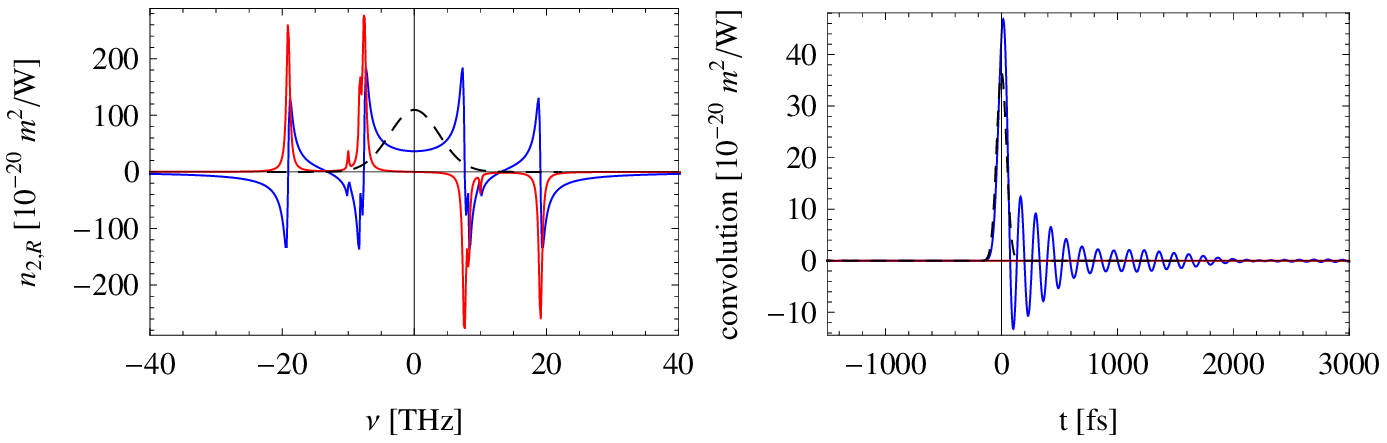}
\includegraphics[width=8.5cm]{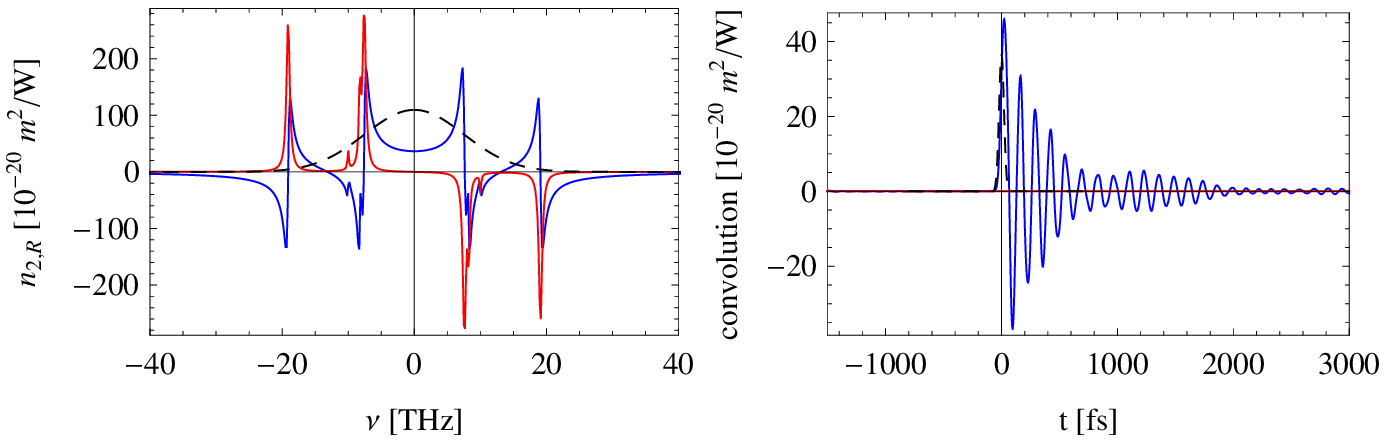}
\caption{\label{fig:convolution} Convolving the Raman response of Fig. \ref{fig:Raman-KJ} with a Gaussian pulse having intensity FWHM of 1000 fs (top), 200 fs (2. plot), 100 fs (3. plot) and 50 fs (bottom). The left plot shows the Raman response, and the Fourier transform of the Gaussian intensity (displayed using arbitrary scaling in the $y$-direction) is shown as a dashed line. The right plot shows the convolved time response (real part as blue and imaginary part as red, the latter always being zero as the pulse is symmetric around zero frequency), and the black dashed line corresponds to the long-pulse limit, where $n^I_{2,R}(t)\rightarrow n^I_{2,R}|A(t)|^2$.
    }
\end{figure}

Presumably the only measurement of the $\chi^{(3)}_{ZZZZ}$ electronic nonlinearity was performed by Wang et al. \cite{wang:2005}, where they used a 0.06\% Fe doped MgO:LN crystal and a 200 fs 1 kHz $Z$-scan setup at 520 nm. Fe-doped crystals are used for enhancing photorefractive effects, but on the other hand on a sub-ps scale with kHz repetition rates they should not kick in. They measured $\chi^{(3)}_{ZZZZ}=4.96\times 10^{-13}$ esu, i.e. $\chi^{(3)}=0.69\times 10^{4}~\rm pm^2/V^2$ or $n^I_{2}=39\times 10^{-20}~\rm m^2/W$. Whether we can assume that a 200 fs pulse does not accumulate any contribution to the nonlinear refractive index is an open question: they claimed that a time-resolved nonlinearity measurement did not reveal any non-instantaneous response, and therefore concluded that the measured nonlinearity is purely instantaneous. We checked this by convolving the Raman response of Fig. \ref{fig:Raman-KJ} with a Gaussian pulse having intensity FWHM from 50-1000 fs, see Fig. \ref{fig:convolution}. What we are interested in is
\begin{align}\label{eq:convolve}
    n^I_{2,R}(t)=n^I_{2,R}\int_{-\infty}^\infty dt' h_R(t-t')|A(t')|^2
\end{align}
where $A$ is the Gaussian pulse and $h_R$ is the normalized Raman response, Eq. (\ref{eq:hRt}). The peak value at $t=0$ will measure the Kerr-like SPM nonlinearity picked up through the Raman interaction. Obviously if $A$ is a long pulse, it will be $\delta$-like in frequency domain, and the peak value will simply be $n^I_{2,R}$, i.e. the DC value of the Raman response. This is what we observe in Fig. \ref{fig:convolution} for 1000 fs and even 200 fs input (top two plots): the convolution is almost identical to the long-pulse limit where $n^I_{2,R}(t)\rightarrow n^I_{2,R}|A(t)|^2$, which is indicated with a black dashed line. When 100 fs or 50 fs pulses are used, the bandwidth of the excitation is large enough to couple into Raman modes, and the ringing in the convolution starts to appear. Note three things: (a) some Kerr-like nonlinearity is "picked" up from the Raman term even with short pulses, and this must be taken into account if short pulses are used for $Z$-scan measurements. (b) The explanation used by Wang et al. that they only observe instantaneous nonlinearities is not quite true. What they measure is the electronic plus DC Raman component, the former being instantaneous and the latter actually some kind of steady state. (c) The fact that they did not observe any delayed response on the nonlinearity was exactly because they did not use a short enough pulse to excite the Raman modes. As before we can correct for the cascading contributions, which at 520 nm turns out to be $n^I_{2,\rm casc}=-13\times 10^{-20}~\rm m^2/W$. Thus we conclude that
\begin{align}\label{eq:n2-LN-wang-corrected}
    n^I_{2,\rm tot}&=52\times 10^{-20}~{\rm m^2/W}
\end{align}

\subsection{Evaluation of the data}
\label{sec:evaluation}

\begin{figure}[tb]
\includegraphics[width=8.5cm]{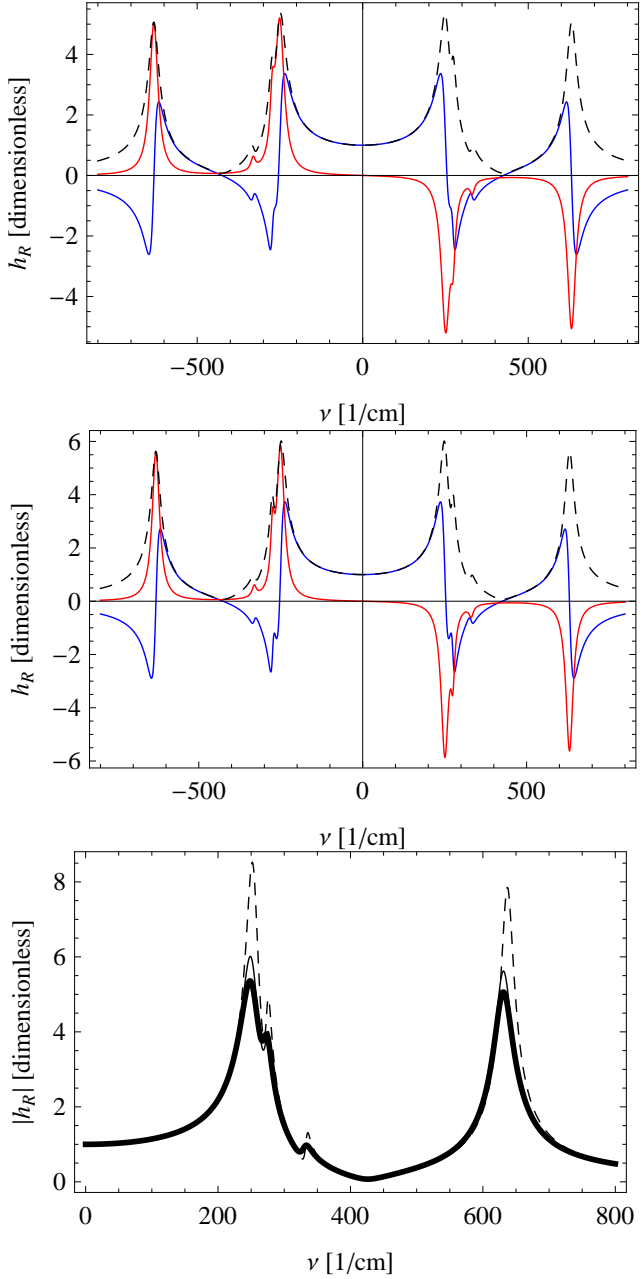}
\caption{\label{fig:Raman-KJ-5MgOcLN} Raman model from Kaminow and Johnston \cite{Kaminow:1967}, i.e. Fig. \ref{fig:Raman-KJ} normalized to unity at DC. We modified the line widths for 5\% MgO:cLN (top) and cLN (middle). The bottom plot compares the absolute values of the original model, presumably sLN (thin dashed) and our modified model for 5\% MgO:cLN (thick) and cLN (thin).
    }
\end{figure}

We choose to use the Raman modes found by Kaminow and Johnston for our simulations. Their experiment was probably carried out using an sLN crystal, see more below. One aspect that is relevant here is that we usually use cLN and often cLN with 5\%MgO doping. The cLN crystals turn out to have broader lines than sLN \cite{Schwarz:1997,Lengyel:2007pssc}, in particular when using the 5\%MgO doped cLN. Also the resonance frequency is shifted slightly \cite{Lengyel:2007pssc}, which is of little importance for our purpose. We therefore use the following line widths and resonance frequencies for a 5\%MgO:cLN at room temperature
\begin{align}\label{eq:raman-MgOcLN-begin}
    \nu_\sigma&=\{251,273,331,631\}\icm\\
    \Gamma_\sigma&=\{32,16,16,30\}\icm\\
    \tau_{1,\sigma}&=\{21.2, 19.5, 16.0, 8.4\}~{\rm fs}\\
    \tau_{2,\sigma}&=\{332, 664, 664, 354\}~{\rm fs} \label{eq:raman-MgOcLN-end}
\end{align}
The third mode line width is a bit unsure as we did not find accurate information about it for this particular crystal, so we assumed it to be twice as big as in sLN. This is of little importance since it is not such a relevant mode. In the same way for cLN we get
\begin{align}\label{eq:raman-cLN}
    \nu_\sigma&=\{251,275,331,631\}\icm\\
    \Gamma_\sigma&=\{28,14,16,27\}\icm\\
    \tau_{1,\sigma}&=\{21.2, 19.3, 16.0, 8.4\}~{\rm fs}\\
    \tau_{2,\sigma}&=\{379, 758, 664, 393\}~{\rm fs}
\end{align}
These choices do not change (substantially) the fractional Raman values, which are given by Eq. (\ref{eq:fRsigma-JK-value}), and we repeat them for convenience
\begin{align}
    f_{R,\sigma}&=\{0.635,0.105,0.020,0.240\}
\end{align}
What does change, though, is the peak-to-DC ratio (clearly seen in Fig. \ref{fig:Raman-KJ-5MgOcLN} bottom): the DC level is found using the analysis above, but when we increase the line width the peak of the Raman response decreases since the area is conserved. At the same time the DC level is unchanged, as discussed before. Thus, the ratio will increase with an increased line width, and we therefore find 
\begin{align}
\label{eq:DC-ratio-KJ-1}
 \frac{\chi_{R}^{(3)}}{{\rm max}[\chi_{R,\sigma}^{(3)}({\rm peak})]}&=0.119, \quad \text{sLN}
\\
\label{eq:DC-ratio-cLN}
 \frac{\chi_{R}^{(3)}}{{\rm max}[\chi_{R,\sigma}^{(3)}({\rm peak})]}&=0.17, \quad \text{cLN}\\
\label{eq:DC-ratio-MgOcLN}
 \frac{\chi_{R}^{(3)}}{{\rm max}[\chi_{R,\sigma}^{(3)}({\rm peak})]}&=0.19, \quad \text{5\%MgO:cLN}
\end{align}
where Eq. (\ref{eq:DC-ratio-KJ-1}) is the value from Eq. (\ref{eq:DC-ratio-KJ}), repeated here for context. The value in Eq. (\ref{eq:DC-ratio-cLN}) is quite similar to the values of Schiek et al. (where a cLN crystal was used) and Phillips et al. (who presumably use cLN crystals). We plot the Raman response function in Fig. \ref{fig:Raman-KJ-5MgOcLN}, and the bottom plot compares the original model (dashed) with the modified models (full).

We see the line widths are crucial for determining the key parameters of the model, and changing them leads to different parameters than those used to determine the coefficients: the gain measurements determine the peak strengths, which in turn through the line widths and resonance frequencies give the DC values for each mode. Whether the cubic nonlinear strength changes with crystal composition is not known, although indications from the measurement of the quadratic nonlinearities indicate that variations are minimal \cite{Shoji:1997}. However, in order to extract the DC parameter $\chi_R^{(3)}$ from a single measurement of a Raman peak gain, we need to relate its value to the DC value using the calculated ratio for that particular Raman model. To do this consistently it is important to know which kind of LN the gain measurement is performed on. Alas, this information is consistently left out. However, educated guesses based on line widths of the lowest mode at $251\icm$ \cite[Fig. 3]{Schwarz:1997} are that (a) Kaminow and Johnston as well as the measurements in Boyd's book use an sLN crystal, because the mode has a line width of $23\icm$ (b) Chunaev et al. use a cLN, because the mode has a line width of $28\icm$. 

There are measurements of Raman gain and $Z$-scan measurements performed at the same (or similar) wavelengths, namely Eq. (\ref{eq:chi3Rsigma-Chuanev}) and Eq. (\ref{eq:chi3Rsigma-Johnston}) for the Raman gain, and Eq. (\ref{eq:n2-LN-desalvo-corrected}) for the $Z$-scan measurement, which gave $n^I_{2,\rm tot}=38.6\pm3.2\times 10^{-20}~{\rm m^2/W}$ at $\lambda=1.064\mic$. We now use this value as a reference point for the total nonlinearity at this wavelength. The problem lies now in determining the DC Raman value from the peak gain values. If we use the ratio 0.17 for cLN, cf. Eq. (\ref{eq:DC-ratio-cLN}), then we get for the Chunaev et al. data at $\lambda=1.047\mic$ \cite{Chunaev2006}
\begin{align}
    \label{eq:chi3-Chunaev-tot-value}
    \chi_{R}^{(3)}&=0.61\pm0.092\times 10^4~\rm pm^2/V^2\\
    \label{eq:n2-Chunaev-tot-value}
    n^I_{2,R}&=37.0\pm 6.4\times 10^{-20}~\rm m^2/W\\
    \label{eq:fR-Chunaev}
    f_{R}&=0.96\pm 0.18
\end{align}
Frequency scaling between $\lambda=1.064\mic$ and $\lambda=1.047\mic$ is considered insignificant. In the uncertainty calculations we estimated the error on the Raman gain measurement to be $\pm15\%$, which is quite conservative. Such a high Raman fraction, close to unity, seems quite unphysical. However, strong nuclear contributions of around 60\% to the Kerr nonlinearity has been observed in niobium oxide glasses \cite{royon:2007}, so the result more than anything is an indication that LN has very strong Raman nonlinearities. We also remark the large uncertainty, and within two standard deviations one would have a value of around $f_R=0.6$, i.e. along the lines what Schiek et al. obtained, cf. Eq. (\ref{eq:fR-Schiek}). If we use a smaller ratio than 0.17 then the $f_R$ value will be smaller.



The measurement by DeSalvo et al. leading to Eq. (\ref{eq:n2-LN-desalvo-corrected-532}), $n^I_{2,\rm tot}=93\times 10^{-20}~{\rm m^2/W}$, lies close in pump wavelength to the Raman measurements of Johnston and Kaminow, and we can therefore make a cautious calculation of the ratio between the total, Eq. (\ref{eq:n2-LN-desalvo-corrected-532}), and the Raman DC nonlinearity, Eq. (\ref{eq:n2-JK-tot-value}), to get
\begin{align}\label{eq:fR-LN-532}
f_R=0.37\pm0.08
\end{align}
This value is therefore a quite good estimate at 532 nm, under the simple assumption of a negligible change in the Raman nonlinearity (based on the scaling discussed above, it should be less than 10\%, and the fraction should therefore be accurate within this range). This value is considerably smaller than the value obtained above, but remember that at 532 nm the 2BM predicts a strong enhancement of the electronic nonlinearity due to onset of two-photon absorption. Since the Raman part does not follow this scaling, the result is a much reduced Raman fraction. 

In Ref. \cite{zhou:2012} we used a 5\%MgO:cLN and found that the following parameters gave a good agreement with the experimental results at $\lambda_1=1.3\mic$
\begin{align}\label{eq:n2-LN-our}
    n^I_{2,\rm tot}&=45\times 10^{-20}~{\rm m^2/W}, \quad f_R=0.50
\end{align}
There we used the 4-line Raman model that we presented in Eqs. (\ref{eq:raman-MgOcLN-begin})-(\ref{eq:raman-MgOcLN-end}). In another recent publication \cite{bache:2011a} we used the same Raman model and parameters but a slightly lower $n^I_{2,\rm tot}$ value. The value $f_R=50\%$ was in part inspired by the analysis presented above, where evidently very large Raman fractions are possible, and in part by looking for agreement between the numerical model and selected experimental data. 

Phillips et al. \cite{Phillips:2011} found a different value at $\lambda=1.043\mic$. They also used $n^I_{2,\rm tot}=38.6\times 10^{-20}~{\rm m^2/W}$, which was estimated in the same way as we did for Eq. (\ref{eq:n2-LN-desalvo-corrected}), and now using the previously calculated value Eq. (\ref{eq:n2-Phillips-tot-value}) we get
\begin{align}\label{eq:fR-LN-Phillips}
f_R=0.14
\end{align}
i.e. a much lower total Raman fraction; this is due to the very low peak Raman susceptibility they chose based on data from simulations and comparison with experiments.  

A final way of scaling the data is to take a single $Z$-scan measurement, and pair it with all the possible Raman data. This will give various results for the electronic value. Now that we know the electronic value we can argue that it scales as either (a) Miller's delta or (b) as the shape of the 2-band model (i.e. we assume that the 2-band model can predict the shape if not the exact value).

\section*{Acknowledgments}

\label{sec:Acknowledgements}

The Danish Council for Independent Research (grants no. 21-04-0506, 274-08-0479, and 11-106702) is acknowledged for support.

\bibliographystyle{apsrev4-1long}
\bibliography{d:/Projects/Bibtex/literature}

\end{document}